\documentclass[12pt,letterpaper]{article}
\usepackage[top=0.85in,left=1in,footskip=0.5in,marginparwidth=2in]{geometry}
\usepackage{color}

\usepackage{comment}
\usepackage{amsmath,amssymb}
\usepackage{multirow}
\usepackage{url}

\usepackage{newtxtext,newtxmath}
\usepackage[T1]{fontenc}

\usepackage[numbers,sort]{natbib}

\usepackage{setspace}
\usepackage{booktabs}
\usepackage{threeparttable}
\usepackage{blkarray}

\usepackage{nameref,hyperref}
\usepackage{cleveref}
\usepackage[right]{lineno}

\usepackage{microtype}
\DisableLigatures[f]{encoding = *, family = * }

\raggedright
\usepackage{changepage}

\usepackage[aboveskip=1pt,labelfont=bf,labelsep=period,singlelinecheck=off]{caption}

\makeatletter
\renewcommand{\@biblabel}[1]{\quad#1.}
\makeatother

\usepackage{lastpage,fancyhdr,graphicx}
\usepackage{epstopdf}
\renewcommand{\footrule}{\hrule height 2pt \vspace{2mm}}
\fancyheadoffset[L]{2.25in}
\fancyfootoffset[L]{2.25in}

\usepackage{color}
\usepackage{soul}

\definecolor{Gray}{gray}{.25}

\usepackage{graphicx}

\usepackage{sidecap}
\usepackage{ragged2e}

\usepackage{wrapfig}
\usepackage[pscoord]{eso-pic}
\usepackage[fulladjust]{marginnote}
\reversemarginpar

\usepackage{verbatim}
\usepackage{algorithm}
\usepackage{algpseudocode}

\newcommand{\Rmnum}[1]{\uppercase\expandafter{\romannumeral #1}}

\newcommand{\algcall}[2]{\textproc{#1}(#2)}

\hypersetup{
hidelinks
}
\newcommand{\keywords}[1]{
\begin{flushleft}
\textbf{Keywords:} #1
\end{flushleft}
}

\begin{document}
\pagestyle{plain}
\pagestyle{myheadings}
\fancyhf{}
\rfoot{\thepage/\pageref{LastPage}}
\renewcommand{\footrule}{\hrule height 2pt \vspace{2mm}}
\fancyheadoffset[L]{2.25in}
\fancyfootoffset[L]{2.25in}
\vspace*{0.35in}

\begin{center}
{\Large
\textbf\newline{A Gene Ranking Framework Enhances the Design Efficiency of Genome-Scale Constraint-Based Metabolic Networks under Time Limits} 
}
\newline
\\
Yier Ma$^{1,2}$, Takeyuki Tamura$^{1,2}$ \\
\end{center}

\begin{flushleft}
$^1$Bioinformatics Center, Institute for Chemical Research,
Kyoto University, Kyoto, Japan \\
$^2$Graduate School of Informatics, Kyoto University, Kyoto, Japan \\
\texttt{Email: mayier@kuicr.kyoto-u.ac.jp, tamura@kuicr.kyoto-u.ac.jp}
\end{flushleft}
\bigskip

\justifying
\section*{Abstract}
The design of genome-scale constraint-based metabolic networks has steadily advanced, with an increasing number of successful cases achieving growth-coupled production, in which the biosynthesis of key metabolites is linked to cell growth. 
However, a major cause of design failures is the inability to find solutions within realistic time limits. Therefore, it is essential to develop methods that achieve a high success rate within the specified computation time. 
In this study, we propose a framework for ranking the importance of individual genes to accelerate the solution of the original mixed-integer linear programming (MILP) problems in the design of constraint-based models. 
In the proposed method, after pre-assigning values to highly important genes, the MILPs are solved in parallel as a series of mutually exclusive subproblems. 
It is found that our framework was able to recover most of the successful cases identified by the original approach and achieved a 37$\%$ to 186$\%$ increase in success rate compared to the original method within the same time limits. 
Analysis of the MILP solution process revealed that the proposed method reduced the sizes of subproblems and decreased the number of nodes in the branch-and-bound tree. 
This framework for ranking gene importance can be directly applicable to a range of MILP-based algorithms for the design of constraint-based metabolic networks.
The developed scripts are available on \href{https://github.com/MetNetComp/Gene-Ranked-RatGene}{https://github.com/MetNetComp/Gene-Ranked-RatGene}.
\keywords{Biochemistry, Constraint optimization, Genetics, Integer linear programming, Metabolic Networks}

\section{Introduction}\label{sec:intro}
Mathematical modeling of metabolism is essential for quantifying key properties of metabolic systems and has greatly contributed to advances in metabolic engineering. It provides a quantitative framework to describe cellular physiology and to estimate the usage of metabolic pathways, enabling a clearer understanding of metabolic behavior. Unlike heuristic approaches, mathematical models can explain complex regulatory mechanisms and generate predictions beyond experimentally tested conditions while ensuring robustness and reproducibility \cite{yasemi2021modelling}.
Metabolic modeling is broadly classified into two approaches: kinetic modeling and constraint-based modeling \cite{maranas2016optimization, nielsen2017systems, yasemi2021modelling}. Kinetic modeling describes dynamic changes in metabolite concentrations using nonlinear ordinary differential equations (ODEs) that incorporate parameters such as enzyme expression \cite{nielsen2017systems}. Although this approach offers detailed insight into metabolic dynamics, it is typically limited to small-scale systems due to high computational demands and the scarcity of reliable experimental data \cite{maranas2016optimization, kadir2010modeling, khodayari2014kinetic}. To overcome these limitations, the non-equilibrium steady-state assumption is often adopted \cite{gombert2000mathematical, fleming2012mass}.
Both experimental evidence and mathematical analyses support the validity of this assumption under constant growth conditions in batch cultures \cite{stephanopoulos1998metabolic, schauer1983quasi, gottstein2016constraint}.

Constraint-based modeling represents metabolism using linear constraints under the steady-state assumption, thereby substantially reducing model complexity \cite{kim2012recent, feist2007genome}. Under this assumption, metabolite concentrations remain constant and the model is mainly determined by reaction stoichiometry. Stoichiometry specifies the quantitative relationships between reactants and products in chemical reactions, indicating how many molecules of each compound participate in a reaction. When expressed in matrix form, these coefficients form the stoichiometric matrix, which encodes the structure of the metabolic network \cite{maranas2016optimization}. Over the past decades, constraint-based models have been developed for a wide range of viral species and compiled in public databases, further expanding their applicability to metabolic network analysis and design \cite{orth2010reconstruction, orth2011comprehensive, norsigian2020bigg}.

Each constraint-based model contains a virtual reaction that represents cellular growth. This growth reaction is constructed to reflect results obtained from biological experiments. In standard simulations of constraint-based models, the cellular growth rate (GR) is maximized, since strains with higher growth rates are more likely to be maintained during serial passaging. In contrast, the reaction responsible for producing the target metabolite is referred to as the production reaction, and its production rate is denoted as PR. Accordingly, the designed strains are assessed based on the PR obtained under the condition of GR maximization in the simulations.

Growth-coupled production refers to the simultaneous occurrence of cell growth and target metabolite production. However, only a limited number of target compounds are naturally growth-coupled. Therefore, appropriate gene deletion strategies are required to achieve growth-coupled production.

Gene–protein–reaction (GPR) associations describe the relationships among genes, proteins, and reactions in metabolic networks. They specify how genes encode enzymes (proteins) that catalyze biochemical reactions. These associations are commonly represented using logical operators such as “AND” and “OR,” where “AND” indicates that multiple genes jointly form an enzyme complex, and “OR” indicates that alternative genes can independently catalyze the same reaction \cite{maranas2016optimization}. By integrating genetic information with metabolic reaction networks, GPR associations enable metabolic network design and analysis at the gene level.

\begin{figure*}[htbp]
    \centering
    \includegraphics[scale=0.4, trim={0cm 0.5cm 0cm 0.5cm}, clip]{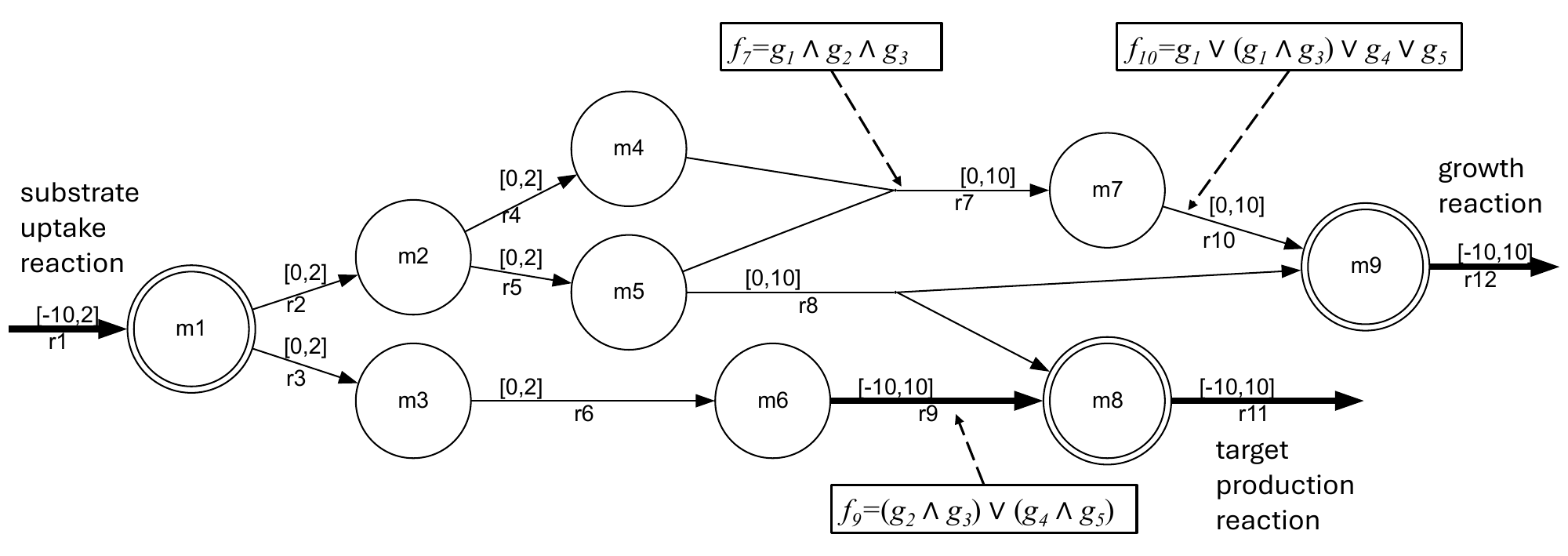}
    \caption{An example of a constraint-based metabolic network. $r_1$ to $r_{12}$ are reactions, $m_1$ to $m_9$ are metabolites and $g_1$ to $g_5$ are genes. The intervals marked next to reactions are the lower and upper bounds for reactions. A negative lower bound indicates a reversible reaction. Three valid GPR associations are embedded in this network. $r_1$ corresponds to the substrate uptake reaction, $r_{12}$ to the growth reaction. $m_8$ is defined as the key metabolite, and $r_{11}$ is the target production reaction for its production.}
    \label{fig1}
\end{figure*}

Mixed-integer linear programming (MILP) plays a key role in the design of constraint-based metabolic networks. Constraint-based models describe biological systems using linear constraints that capture physiological limitations \cite{orth2010flux, varma1994metabolic}. MILP enables the identification of optimal solutions while satisfying all imposed constraints. These problems typically involve both continuous variables, such as reaction fluxes, and discrete variables, such as gene activation or repression \cite{burgard2003optknock}.
By handling continuous and discrete variables simultaneously, MILP allows the modeling of complex decision structures. Logical (Boolean) relationships commonly arise in metabolic network design and can be expressed as integer constraints, making MILP well-suited for hybrid systems that combine logical rules with continuous dynamics \cite{tamura2023gene}. In addition, many design problems require the optimization of multiple objectives \cite{ranganathan2010optforce, zomorrodi2012optcom}. MILP supports multi-objective formulations, enabling systematic exploration of trade-offs within the feasible solution space.

Although MILP is highly useful for the design of constraint-based metabolic networks, MILP problems are NP-complete. As a result, even when a problem can be formulated as an MILP, it often cannot be solved within practical time limits. To address this issue, it is necessary to develop MILP-based frameworks for metabolic network design that can identify feasible and effective design strategies within a given time constraint.

In this study, we propose an acceleration method for solving MILP problems by preferentially assigning binary values (0 or 1) to genes identified as important based on GPR associations and network topological information. During the MILP solving process, these important genes are fixed at an early stage to guide the search efficiently. Computational experiments were performed using representative constraint-based models of Escherichia coli (e$\_$coli$\_$core, iML1515, iJO1366, and iAF1260) and the yeast model iMM904. The results show that the proposed method increases the number of successful cases in which feasible solutions are found within a fixed time limit. Further analysis of the MILP solving process shows that pre-assigning values to important genes reduces the computational complexity of the MILP problems.

\section{Preliminaries}\label{sec:prelim}
\subsection{Constraint-based Metabolic Network}
A constraint-based metabolic network is defined as $N = \{R, M, S, G, F, LB, UB\}$ consisting of three fundamental components and their interactions, including \textbf{reactions} $R$, \textbf{metabolites} $M$, and \textbf{genes} $G$. In this context, $S$ and $F$ represent a \textbf{stoichiometric matrix} and a set of \textbf{GPR} associations, respectively. $LB$ and $UB$ correspond to sets of \textbf{lower bounds} and \textbf{upper bounds} for rates of reactions $R$. The set of reactions $R$ is categorized into reversible reactions and irreversible reactions. The stoichiometric information of the network is encoded within $S$ and the size of $S$ is $m \times n$, where $m$ and $n$ are the number of metabolites and reactions, respectively. Each entity $s_{j,i}$ of the matrix represents the coefficient of metabolite $m_j$ in reaction $r_i$. A positive $s_{j,i}$ indicates $m_j$ is produced by $r_i$, and a negative $s_{j,i}$ corresponds to consumption of $m_j$, whereas a zero value denotes $m_j$ is not involved in $r_i$. The biomass reaction $r_{biomass}$ is the reaction that produces all necessary metabolites synthesized for cell growth. A target production reaction $r_{target}$ is a reaction that produces the target metabolite.

An example network is shown in Fig. \ref{fig1}. $\{m_1, m_2, \dots, m_9\}$ are five metabolites and $\{r_1, r_2, \dots, r_{12} \}$ are twelve reactions. $r_1$, $r_9$, $r_{11}$, and $r_{12}$ are reversible reactions that can proceed in both directions, and the rest are irreversible reactions. The intervals on reactions are the lower and upper bounds for reaction rates. A negative lower bound indicates a reversible reaction. Here, $r_1$ corresponds to the substrate uptake reaction, $m_8$ to the target metabolite, and $m_9$ to the metabolite for cell growth. Thus, the reaction $r_{11}$ transfers the target metabolite $m_8$ is the target production reaction, and $r_{12}$ transfers $m_9$ is the biomass reaction. The coefficients for each metabolite in all reactions are $1$ except for $m_7$ in $r_7$. The coefficient of $m_7$ in $r_7$ is $3$. Based on the above definitions, the stoichiometric matrix $S$ for the example network in Fig. \ref{fig1} is constructed as:
\begin{equation}
\scalebox{0.9}{
\begin{blockarray}{ccccccccccccc}
& $r_1$ & $r_2$ & $r_3$ & $r_4$ & $r_5$ & $r_6$ & $r_7$ & $r_8$ & $r_9$ & $r_{10}$ & $r_{11}$ & $r_{12}$ \\
\begin{block}{c(cccccccccccc)}
$m_1$ & 1 & -1 & -1 & 0 & 0 & 0 & 0 & 0 & 0 & 0 & 0 & 0 \\ 
$m_2$ & 0 & 1 & 0 & -1 & -1 & 0 & 0 & 0 & 0 & 0 & 0 & 0 \\ 
$m_3$ & 0 & 0 & 1 & 0 & 0 & -1 & 0 & 0 & 0 & 0 & 0 & 0 \\ 
$m_4$ & 0 & 0 & 0 & 1 & 0 & 0 & -1 & 0 & 0 & 0 & 0 & 0 \\ 
$m_5$ & 0 & 0 & 0 & 0 & 1 & 0 & -1 & -1 & 0 & 0 & 0 & 0 \\ 
$m_6$ & 0 & 0 & 0 & 0 & 0 & 1 & 0 & 0 & -1 & 0 & 0 & 0 \\ 
$m_7$ & 0 & 0 & 0 & 0 & 0 & 0 & 3 & 0 & 0 & -1 & 0 & 0 \\ 
$m_8$ & 0 & 0 & 0 & 0 & 0 & 0 & 0 & 1 & 1 & 0 & -1 & 0 \\ 
$m_9$ & 0 & 0 & 0 & 0 & 0 & 0 & 0 & 1 & 0 & 1 & 0 & -1 \\
\end{block}
\end{blockarray} \nonumber
}
\end{equation}
Each row represents the relationship between a single metabolite and all reactions. For example, the second row indicates that metabolite $m_2$ is generated by reaction $r_2$ but consumed by reactions $r_4$ and $r_5$. Each column represents the relationship between a reaction and all the network's metabolites. For instance, reaction $r_3$ is shown to produce metabolites $m_3$ from the consumption of metabolite $m_1$ in the third column.

\subsection{Growth-coupled Production}
The steady-state assumption refers to a condition in which the concentration of a metabolite is dynamically balanced over time. For example, a flux distribution $[2,1,1,0,0,0,0,0,0,0,0,0]^T$ confirms that the change in concentration of $m_1$ is zero in the example network $N$ because the inner product of this vector and the first row of matrix $S$ results $0$. The steady-state assumption of the whole metabolic network ensures that concentrations of all metabolites are stable. $S \cdot x =0$ formulates this steady-state constraint. $x$ is a vector of variables that correspond to the rates of all reactions.

Growth-coupled production has two most common types by definition: strong-coupled production and weak-coupled production. The strong-coupled production is that the production of the target metabolite is strictly possible in all non-zero fluxes with substrate uptake. The weak-coupled production means that the production of the target metabolite is possible in all fluxes with the maximal biomass reaction rate\cite{schneider2021systematizing}. In this context, we focus on weak-coupled production.

Let $x_i$ be the continuous variable representing the rates of reaction $r_i$. In particular, $x_{biomass}$ and $x_{target}$ are the two variables that represent rates of biomass reaction and target production reaction, respectively. Let $y_k$ be the binary variable indicating the existence of the gene $g_k$. A knockout strategy $K \in G$ is a set of genes. $K$ satisfies \{ $y_{\gamma} = 0 \, | \, \gamma \in K$ \} and \{ $y_{\gamma} = 1 \, | \, \gamma \notin K$ \}. Form the following linear programming (LP) problem $P_1$ based on $K$:
\begin{align}
    & max \quad x_{biomass} \\ 
    & s.t. \quad S \cdot x = 0 \nonumber \\
    & \quad \quad p_i \cdot lb_i \leq x_i \leq p_i \cdot ub_i \nonumber  \\
    & \quad \quad p_i = f_i(G) \nonumber 
\end{align}
Let $v_{biomass}$ denote the solution of problem $P_1$, which is the optimal value of the biomass reaction rate as well. Then construct the following LP problem $P_2$:
\begin{align}
    & min \quad x_{target} \\ 
    & s.t. \quad S \cdot x = 0 \nonumber \\
    & \quad \quad p_i \cdot lb_i \leq x_i \leq p_i \cdot ub_i, \quad i \neq biomass \nonumber  \\
    & \quad \quad  x_{biomass} = v_{biomass} \nonumber  \\
    & \quad \quad p_i = f_i(G) \nonumber
\end{align}
Let $v_{target}$ denote the solution of problem $P_2$. If both $v_{biomass}$ and $v_{target}$ are greater than their individual preset thresholds, we call that the deletion strategy $K$ achieves growth-coupled production.  

\begin{table}[htbp]
\footnotesize
\captionsetup{justification=centering}
\caption{Deletion strategies and Flux distributions}\label{tf}
\centering
\begin{tabular}{@{}ccc@{}c@{}c@{}c@{}c@{}c@{}c@{}c@{}c@{}c|@{}c|@{}c@{}}
\hline
Deletions & Problem & $v_1$ & $v_2 $& $v_3$ & $v_4$ & $v_5$ & $v_6$ & $v_7$ & $v_8$ & $v_9$ & $v_{10}$ & $v_{11}$ & $v_{12}$ \\
\hline
$\varnothing$ & $P_1(flux)$ & 2 & 2 & 0 & 1 & 1 & 0 & 1 & 0 & 0 & 3 & 0 & 3 \\
$\varnothing$ & $P_2(flux)$ & 2 & 2 & 0 & 1 & 1 & 0 & 1 & 0 & 0 & 3 & 0 & 3 \\
\hline
$g_4,g_5$ & $P_1(flux)$ & 2 & 2 & 0 & 1 & 1 & 0 & 1 & 0 & 0 & 3 & 0 & 3 \\
$g_4,g_5$ & $P_2(flux)$ & 2 & 2 & 0 & 1 & 1 & 0 & 1 & 0 & 0 & 3 & 0 & 3\\
\hline
$g_1$ & $P_1(flux)$ & 2 & 2 & 0 & 0 & 2 & 0 & 0 & 2 & 0 & 0 & 2 & 2  \\
$g_1$ & $P_2(flux)$ & 2 & 2 & 0 & 0 & 2 & 0 & 0 & 2 & 0 & 0 & 2 & 2  \\
\hline
$g_1,g_4,g_5$ & $P_1(flux)$ & 2 & 2 & 0 & 0 & 2 & 0 & 0 & 2 & 0 & 0 & 2 & 2 \\
$g_1,g_4,g_5$ & $P_2(flux)$ & 2 & 2 & 0 & 0 & 2 & 0 & 0 & 2 & 0 & 0 & 2 & 2 \\
\hline
\end{tabular}
\end{table}

As a further illustration, we again consider the example of $N$ in Fig. \ref{fig1}. The thresholds for production rates of target metabolite $m_8$ and biomass $m_9$ are preset at $1$ and $1$, respectively. Table~\ref{tf} shows the flux distributions obtained from the above problem $P_1$ and $P_2$ under different gene deletion strategies. When the genes are not deleted, or when genes $g_4$ and $g_5$ are deleted, the flux distribution reveals that the maximum rate of the biomass reaction $v_{12}$ is $3$. By constraining the rate of this reaction to its maximum value, the flux distribution shows that the value of the target production rate $v_{11}$ is $0$. This is the growth-coupled rate for the target production reaction in this case. Although the biomass reaction rate is greater than the required threshold, the target production reaction rate is unable to achieve the preset threshold. Consequently, the growth-coupled production is not satisfied in such conditions. When the deletion strategies $g_1$ and $\{g_1,g_4,g_5\}$ are adopted, maximal value for $v_{12}$ is $2$. Then the growth-coupled production rate $v_{11}$ for the target metabolite is $2$. Both reaction rates satisfy the thresholds. Thus, the growth-coupled production is achieved under the condition of such deletion strategies.

\section{Method}\label{sec:method}
\subsection{Gene Ranking Strategies}
To evaluate the importance of genes based on prior information, five distinct strategies for scoring genes on the basis of GPR associations and topological structures are proposed in this study. GPR associations represent fundamental knowledge for describing the associations between genes and metabolic reactions. These relationships are typically formulated as Boolean functions, and they cover the combinatorial logic by which gene products, primarily enzymes encoded, govern reaction activities within metabolic networks. Such formulations not only enable integrating genome information into metabolic networks but also provide a basis for assessing the individual contributions of each gene to the network. Table~\ref{t:math} provides an overview of all five strategies, which will be discussed in detail from subsection \ref{sub:start} to \ref{sub:end}.

\vspace{-3mm}
\begin{table}[htbp]
\footnotesize
\captionsetup{justification=centering}
\caption{Gene Ranking strategies Summary}\label{t:math}
\centering
\begin{tabular}{@{}c@{}c@{}c@{}c@{}}
\hline
ID & Name & Definition & Eq. ID\\
\hline
St1 & multiplicity & $m_N(g_k)/|F|$ & (\ref{e:m}) \\
St2 & frequency & $\sum_{i=1}^{n}1_{U(f_i)}(g_k)/|F|$& (\ref{e:f})  \\
St3 & logic & $\sum_{i=1}^{n}Score_{logic}^{f_i}(g_k)$& (\ref{e:l})   \\
St4 & degree & $\sum_{i=1}^{n}Score_{logic}^{f_i}(g_k)  deg(r_i)$& (\ref{e:de})  \\
St5 & revdegree & $\sum_{i=1}^{n}Score_{logic}^{f_i}(g_k)  deg(r_i) rev(r_i)$& (\ref{e:rde})  \\ 
\hline
\end{tabular}
\end{table}

\subsubsection{Strategy of Multiplicity}\label{sub:start}
We firstly introduce a metric of gene importance derived from the multiplicity of a gene appearing across all GPR associations in a network, and introduce a quantitative way to evaluate this metric. To better illustrate definitions, the example $N$ in Fig.\ref{fig1} is taken here. $N$ is a metabolic network with five genes and three GPR associations: $f_7=g_1 \wedge g_2 \wedge g_3$, $f_9=(g_2 \wedge g_3) \vee (g_4 \wedge g_5)$, and $f_{10}=g_1 \vee (g_1 \wedge g_3) \vee g_4 \vee g_5$. Define a multi-set $U(f_i)=\{ g_t \, | \, g_t \in f_i \}$ for a GPR association as a multi-set including a finite number of genes which appear in $f_i$. Then, we have: 
\begin{align}
& U(f_7)=\{ g_1, g_2, g_3 \} \nonumber \\
& U(f_9)=\{ g_2, g_3, g_4, g_5 \} \nonumber \\
& U(f_{10})=\{ g_1, g_1, g_3, g_4, g_5 \} \nonumber  
\end{align}
Define $SQ(f_i)=\{f_i^1, \, f_i^2, \, ..., \, f_i^{\epsilon} \}$ as the multi-set of sequence for a GPR association $f_i$. $f_i^{\epsilon}$ is the identity number of the $\epsilon$-th gene that appears in $f_i$. Then, we have:
\begin{align}
& SQ(f_7)=\{ 1, 2, 3 \} \nonumber \\
& SQ(f_9)=\{ 2, 3, 4, 5 \} \nonumber \\
& SQ(f_{10})=\{ 1, 1, 3, 4, 5 \} \nonumber 
\end{align}
Define the frequency of occurrence of a gene $g_k$ in a multi-set $U(f_i)$ for a GPR association as:
\begin{align}
    & m_{U(f_i)}(g_k) \, := \, \# \{ t \in SQ(f_i) \, | \, g_t = g_k\}
\end{align}
$g_1$ appears once and twice in $f_7$ and $f_{10}$, respectively. But it does not occur in $f_9$. The $m_U$ for the example is:
\begin{align}
& m_{U(f_7)}(g_1)=1 \nonumber \\
& m_{U(f_9)}(g_1)=0 \nonumber \\
& m_{U(f_{10})}(g_1)=2 \nonumber
\end{align}
Therefore, the total multiplicity of a gene $g_k$ in a metabolic network $N$ is defined as:
\begin{align}
    & m_N(g_k) := \sum_{i=1}^{n}m_{U(f_i)}(g_k)
\end{align}
The $m_N$ for the example is:
\begin{align}
& m_{N}(g_1)=m_{U(f_7)}(g_1)+m_{U(f_9)}(g_1)+m_{U(f_{10})}(g_1) \nonumber
\end{align}
And the final definition of the score for gene importance in the first strategy $Score_{multiplicity}$ is provided as:
\begin{align}
    & Score_{multiplicity}(g_k) :=  m_N(g_k)/|F| \label{e:m}
\end{align}
where $|F|$ denotes the total number of GPR associations in a metabolic network. $|F|=3$ for the example. And $m_{N}(g_1)=3$. Finally, $Score_{multiplicity}(g_1)=3/3=1$ for $g_1$ in the example.

\subsubsection{Strategy of Frequency}
Similarly, define the second strategy based on the frequency of a gene. We measure whether any gene exists in a given GPR rule. Firstly, define an indicator function $1_{S}(x):S \rightarrow \{0,1\}$ as:
\begin{align}
    & 1_{U(f_i)}(g_k) := \begin{cases} 1 & if \, g_k \in U(f_i) \\ 0 & if \, g_k \notin U(f_i)   \end{cases} \nonumber
\end{align}
$S$ is a multi-set that might include $x$. Subsequently, we quantify the number of GPR associations in which a gene $g_k$ appears and determine the second strategy for gene importance $Score_{frequency}$ as:
\begin{align}
    & Score_{frequency}(g_k) := \sum_{i=1}^{n}1_{U(f_i)}(g_k)/|F| \label{e:f}
\end{align}

As a further illustration, we again consider the example of $N$. Since $g_1$ only exists in $f_7$ and $f_{10}$, we have: 
\begin{align}
& 1_{U(f_7)}(g_1)=1 \nonumber \\
& 1_{U(f_9)}(g_1)=0 \nonumber \\
& 1_{U(f_{10})}(g_1)=1 \nonumber \\
& |F|=3 \nonumber
\end{align}
Then $Score_{frequency}(g_1)=2/3$ for $g_1$.

\subsubsection{Strategy of Boolean Logic}
The two strategies discussed so far regard the genes associated with each GPR rule merely as a simple set, thereby neglecting the Boolean logic embedded in the GPR representation. To address this limitation, we propose a third strategy for gene importance that explicitly incorporates the underlying Boolean relationships. Assign a basis value $\beta$ equally to each GPR rule. For a GPR rule $f_i$ with an AND connection, the evaluation of the entire expression becomes 0 if any individual clause is 0. As a consequence, each clause is regarded as having an equal impact on the overall GPR rule $f_i$. Similarly, for clauses connected by OR relationships within a GPR rule $f_i$, $f_i=0$ only holds in the case when all the clauses are 0. Thus, all clauses combined could affect the entire $f_i$ and share the value $\beta$. Denote the number of clauses in a OR connection as $\lambda_i$. For $f_i$ controlled by a single gene, such a gene will share the whole $\beta$:
\begin{align}
    & Score_{logic}^{f_i}(C_h) := \beta \quad if \, f_i = \bigwedge_{h=1}^{\lambda_i} C_h \label{e:2} \\
    & Score_{logic}^{f_i}(C_h) := \frac{\beta}{\lambda_i} \quad if \, f_i = \bigvee_{h=1}^{\lambda_i} C_h \\
    & Score_{logic}^{f_i}(g_k) := \beta \quad if \, f_i = g_k \label{e:3}
\end{align}
However, $f_i$ in a metabolic network exhibits hierarchical nesting of multiple clauses, resulting in more complex logical structures for literals in clauses. It is not obvious to derive gene importance from primary clauses. Therefore, \algcall{RecurScore}{$f_i$, $g_k$, $\beta$} function in (\ref{algo1}) is developed to calculate $Score_{logic}^{f_i}(g_k)$ for $g_k$ recursively. $Score_{logic}^{f_i}(g_k)$ is defined as the importance score of a gene $g_k$ in a GPR rule $f_i$ given a basis value $\beta$ based on the above principles (\ref{e:2}) to (\ref{e:3}). The third strategy for determining the importance of a gene is to sum all $Score_{logic}^{f_i}(g_k)$ in the entire metabolic network as:
\begin{align}
     Score_{logic}(g_k) &:= \sum_{i=1}^{n}Score_{logic}^{f_i}(g_k) \nonumber \\ 
    &= \sum_{i=1}^{n}\algcall{RecurScore}{f_i, g_k, \beta} \label{e:l}
\end{align}
\begin{align}
    &\algcall{RecurScore}{f_i, g_k, \beta} := \nonumber \\
    &\begin{cases}
    0 & if \, g_k \notin U(f_i), \\
    \beta & if \, f_i = g_k, \\
    \sum_{h=1}^{\lambda_i}\algcall{RecurScore}{C_h, g_k, \beta} & if \,  f_i = \bigwedge_{h=1}^{\lambda_i} C_h \\
    \sum_{h=1}^{\lambda_i}\algcall{RecurScore}{C_h, g_k, \frac{\beta}{\lambda_i}} & if \,  f_i = \bigvee_{h=1}^{\lambda_i} C_h 
    \end{cases} \label{algo1}
\end{align}
where $\lambda_i$ is the number of clauses in $f_i$.

Take $g_1$ in the case of $N$ as an example. Assume a basis value is $1$. $g_1$ contributes equally compared as to other clauses in $f_7=g_1 \wedge g_2 \wedge g_3$, thus:
\begin{align}
& Score_{logic}^{f_7}(g_1)=1 \nonumber
\end{align}
$f_{10}=g_1 \vee (g_1 \wedge g_3) \vee g_4 \vee g_5$ is a OR connection. The first two clauses that include $g_1$ contribute equally to $f_{10}$. Therefore, we have:
\begin{align}
& Score_{logic}^{f_{10}}(C_1)=1/4 \quad C_1 = g_1\nonumber \\
& Score_{logic}^{f_{10}}(C_2)=1/4 \quad C_2 = g_1 \wedge g_3 \nonumber
\end{align}
$g_1$ monopolizes all weights in $C_1$ and shares the same weight as other literals in $C_2$ which is $(g_1 \wedge g_3)$. Then we have: 
\begin{align}
& Score_{logic}^{C_1}(g_1)=1/4 \nonumber \\
& Score_{logic}^{C_2}(g_1)=1/4 \nonumber \\
& Score_{logic}^{f_{10}}(g_1)=Score_{logic}^{C_1}(g_1)+Score_{logic}^{C_2}(g_1)=1/2 \nonumber
\end{align}
$g_1$ does not exist in $f_9$, thus $Score_{logic}^{f_9}(g_1)=0$. Finally $Score_{logic}$ of $g_1$ should be:
\begin{align}
Score_{logic}(g_1) =& Score_{logic}^{f_7}(g_1)+Score_{logic}^{f_9}(g_1) \nonumber \\ &+Score_{logic}^{f_{10}}(g_1)=3/2 \nonumber
\end{align}

\subsubsection{Strategy of Degree}
The fourth strategy accounts for the degree of each reaction node. The higher degree of a reaction node indicates a larger set of potential metabolic pathways involved. Consequently, the perturbation to such a node may compromise the integrity of the network, thereby indicating its importance within the network. \\
Define in-degree, out-degree, and degree of reaction $r_i$ as:
\begin{align}
    & deg^+(r_i):=|\{ s_{i,j} \, | \, s_{i,j} < 0, \, i=1,2,...,q \}| \nonumber \\
    & deg^-(r_i):=|\{ s_{i,j} \, | \, s_{i,j} > 0, \, i=1,2,...,q \}| \nonumber \\
    & deg(r_i):=deg^+(r_i)+deg^-(r_i)
\end{align}
where $s_{i,j}$ is the entity of the stoichiometric matrix $S$. Then the sum of weighted scores of $Score_{degree}$ is defined as the gene importance for $g_k$:
\begin{align}
    & Score_{degree}(g_k) :=\sum_{i=1}^{n}Score_{logic}^{f_i}(g_k) \cdot deg(r_i) \label{e:de}
\end{align}
We now return to the example of $N$ previously discussed. $Score_{logic}^{f_7}(g_1)=1$, $Score_{logic}^{f_9}(g_1)=0$, and $Score_{logic}^{f_{10}}(g_1)=1/2$. $f_7$, $f_9$, and $f_{10}$ correspond to $r_7$, $r_9$, and $r_{10}$ in Fig. \ref{fig1}, respectively. Then we have: 
\begin{align}
& deg(r_7)=3 \nonumber \\
& deg(r_9)=2 \nonumber \\
& deg(r_{10})=2 \nonumber 
\end{align}
It is derived that:
\begin{align}
& Score_{degree}^{f_7}(g_1)=3 \nonumber \\
& Score_{degree}^{f_9}(g_1)=0 \nonumber \\
& Score_{degree}^{f_{10}}(g_1)=1 \nonumber 
\end{align}
Finally, $Score_{degree}(g_1)=4$ for $g_1$.

\subsubsection{Strategies of Combined Reversibility with Degree}\label{sub:end}
As illustrated in the Fig. \ref{fig1}, reactions are classified into two categories according to their reversibility. Reversible reactions, which can proceed in both forward and backward directions, are associated with a larger number of potential metabolic pathways compared with irreversible ones. They are therefore considered to be of greater importance. Based on this distinction, we introduce the following definitions:
\begin{align}
    & rev(r_i) := \begin{cases} 1 & if \, r_i \in irreversible \; reactions \\ 2 & if \, r_i \in reversible \; reactions   \end{cases} \label{e:rde}
\end{align}
In the case of $N$, reaction $r_7$ and $r_{10}$ are irreversible reactions, while $r_9$ is a reversible reaction. Then it could be derived: 
\begin{align}
& rev(r_7)=1 \nonumber \\
& rev(r_9)=2 \nonumber \\
& rev(r_{10})=1 \nonumber
\end{align}

\subsection{Baseline Algorithm} 
RatGene\cite{ma2025ratgene} is an algorithm that identifies gene deletion strategies for achieving growth-coupled production based on MILP, using the growth-to-production ratio. In the computational experiments of this study, we apply the proposed gene ranking framework to the MILP problems constructed by RatGene. By fixing the values of binary variables corresponding to the most important genes, the original problem is decomposed into multiple subproblems.

Assuming that a set $D$ of $\kappa$ genes is selected, we can construct $2^{\kappa}$ subproblems by assigning values of 0 and 1 to each variable separately. Let $\xi$ be a natural number such that $\xi=1,2,...,2^{\kappa}$. Let us define the $\xi$-th subproblem $P_{\xi}$ as:
\allowdisplaybreaks
\begin{align}
    &min \quad -x_{biomass} + TMGR \cdot \|x_Q\|_0 \\
    &s.t. \, S \cdot x=0 \nonumber \\
    &\,\quad p_i \cdot lb_i \leq x_i \leq p_i \cdot ub_i \nonumber \\
    &\,\quad p_i=f_i(y) \nonumber \\
    &\,\quad x_{biomass} \geq lb^{min}_{biomass} \nonumber \\
    &\,\quad v_{target}/v_{biomass}=\alpha \nonumber \\
    &\,\quad 0 \leq \alpha \leq TMPR/lb^{min}_{biomass} \nonumber \\
    &\,\quad Q:=\{i \ | \ \exists f_i\} \nonumber \\
    &\,\quad y_{d_k}=\lfloor (\xi-1)/2^{\kappa-1} \rfloor \; mod \; 2, \; d_k \in D, \, k=1,2,...,\kappa \nonumber
\end{align}
where the objective function is constructed as minimizing the sum of $l_0$-Norm of the reactions scaled by the theoretical maximum growth rate (TMGR) and the negative biomass reaction rate. The coefficient TMGR is used in the objective function to prioritize the minimization of the $l_0$ norm, and then to maximize the growth rate when the $l_0$ norm is the same. In RatGene, the appropriate $\alpha$ is obtained by iteratively assigning different values to form a series of MILPs. Each binary variable $y_{d_k}$, corresponding to gene $d_k$ in the selected set $D$, is assigned a value of $0$ or $1$. In the MILP problem $P_{RatGene}$ formulated by RatGene, it is possible to decompose the problem into $2^{\kappa}$ subproblems. Therefore, we parallelize the computation of these subproblems.

\begin{table*}[htbp]
\captionsetup{justification=centering}
\caption{Performance comparison on iMM904 dataset}
\label{t1}
\centering
\scriptsize
\begin{threeparttable}
\begin{tabular}{cccccccc}
\toprule
 \multicolumn{2}{c}{Dataset iMM904} & \multicolumn{2}{c}{(1)Only Method \Rmnum{1} Succeeded} & \multicolumn{2}{c}{(2)Both Methods Succeeded} & \multicolumn{2}{c}{(3)Only Method \Rmnum{2} Succeeded}\\
\cmidrule(lr){1-2} \cmidrule(lr){3-4} \cmidrule(lr){5-6} \cmidrule(lr){7-8}
Size \tnote{a} & Method \Rmnum{1} \tnote{b} & Succ. Case & Avg. Time & Succ. Case & Avg. Time & Succ. Case & Avg. Time \\
\midrule
\multirow{2}{*}{1}
& St1,St2,St4 & 50 & 371.28 & 82 & 194.57 & 13 & 386.79 \\
& St3,St5  & 35 & 402.56 & 77 & 175.80 & 18 & 342.60 \\
\midrule
\multirow{4}{*}{2}
& St1,St2 & 108 & 343.02 & 87 & 179.62 & 8 & 276.11 \\
& St3  & 65 & 427.48 & 82 & 191.22 & 13 & 316.76 \\
& St4 & 86 & 417.86 & 82 & 198.57 & 13 & 347.70 \\
& St5 & 66 & 382.55 & 84 & 183.82 & 11 & 304.20\\
\midrule
\multirow{3}{*}{3}
& St1,St2 & 177 & 322.64 & 90 & 157.29 & 5 & 279.17 \\
& St3,St4 & 112 & 379.86 & 86 & 197.36 & 9 & 337.96 \\
& St5 & 151 & 335.54 & 84 & 191.61 & 11 & 339.98 \\
\midrule
\bottomrule
& Method \Rmnum{2} & 95 & 210.92  & 95 & 210.92  & 95 & 210.92 \\
\midrule
\end{tabular}
\begin{tablenotes}
\tiny
\item[a] The number of important genes selected for Method \Rmnum{1}.
\item[b] The time limit for each metabolite is 500 seconds.
\end{tablenotes}
\end{threeparttable}
\end{table*}

\section{Computational Experiments}\label{sec:results}
We conducted computational experiments on three datasets from the BiGG database\cite{norsigian2020bigg} to evaluate the effectiveness of the proposed strategies. The datasets iML1515, iJO1366, iAF1260 and iMM904 represent the genome-scale metabolic networks of \textit{E. coli} or \textit{S. cerevisiae}. The e$\_$coli$\_$core dataset represents a small-scale \textit{E. coli} network. Table~\ref{tab:gem_summary} summarizes information about these models. The Theoretical Maximum Production Rate (TMPR) represents the maximum achievable production rate of the target compound. When TMPR is zero, no gene deletion strategy can achieve growth-coupled production. Therefore, experiments were conducted only for the target compounds with
TMPR$>0$.

\begin{table}[htbp]
\scriptsize
\centering
\caption{Summary of Genome-scale Metabolic Models Used in the Computational Experiments}
\begin{tabular}{lcccc}
\toprule
Model & Reactions & Metabolites & Genes & TMPR$>$0 \\
\midrule
e\_coli\_core & 95   & 72   & 137  & 52  \\
iMM904        & 1577 & 1226 & 905  & 782 \\
iML1515       & 2712 & 1877 & 1516 & 1092 \\
iJO1366       & 2583 & 1805 & 1367 & 1052 \\
iAF1260       & 2382 & 1668 & 1261 & 961 \\
\bottomrule
\end{tabular}
\label{tab:gem_summary}
\end{table}

All experiments were performed on an Ubuntu 20.04 system equipped with an AMD Ryzen Threadripper3 3970X CPU (3.70 GHz, 32 cores / 64 threads). The computational environment included IBM ILOG CPLEX 12.10, the COBRA Toolbox v3.0 \cite{heirendt2019creation}, and MATLAB R2019b. 

\textbf{Method \Rmnum{1}} denotes the proposed gene-ranking-based MILP framework in this study, whereas \textbf{Method \Rmnum{2}} denotes the MILP without the gene-ranking framework, where the original MILPs are constructed by RatGene in the following experiments. All parameters remained the same except for the selected genes when comparing Method \Rmnum{1} and Method \Rmnum{2}. Two main metrics were used for evaluation: the number of successful cases and the average runtime. A successful case is defined as one in which the computational method can successfully identify a valid gene knockout strategy enabling growth-coupled production for a target metabolite.

To ensure a fair and consistent comparison between Method \Rmnum{1} and Method \Rmnum{2}, all analyses were conducted under standardized conditions. Both methods were applied to identical datasets, following the same preprocessing procedures, parameter configurations, and evaluation metrics. This rigorous and consistent evaluation provides a reliable basis for comparing the effectiveness and robustness of the two approaches. In addition, according to the definition of RatGene, the number of iterations corresponds to the number of uniformly sampled points in the ratio constraint space\cite{ma2025ratgene}. Increasing the number of samples improves the likelihood of identifying a feasible solution, but also leads to higher computational costs. To achieve a trade-off between solution feasibility and computational efficiency, the number of iterations is set to 200.

\subsection{Performance Comparison between Methods I and II}
Table~\ref{t1} summarizes the performance of Method \Rmnum{1} using three different sizes of ranked gene sets and Method \Rmnum{2} used as the benchmark on the iMM904 dataset under a fixed time limit. The results of metabolites are divided into three categories: (1) only Method \Rmnum{1} succeeded, (2) both Methods succeeded, and (3) only Method \Rmnum{2} succeeded. As shown in Table~\ref{t1}, Method \Rmnum{1} incorporating gene importance rankings consistently outperformed Method \Rmnum{2} in terms of both the number of successful cases and average runtime. When the most important single gene was selected and assigned 0 and 1, treating each as a mutually exclusive parallel process, Method \Rmnum{1} achieved 36.84$\%$–52.63$\%$ more successful cases than Method \Rmnum{2}, indicating a notable improvement. It also reproduced over 80$\%$ of success cases of Method \Rmnum{2} while reducing runtime, demonstrating both reliability and efficiency. Similarly, when the two most important genes were selected, Method \Rmnum{1} yielded up to 113.68$\%$ additional successful cases, while still recovering approximately 90$\%$ of Method \Rmnum{2}’s results and reducing runtime. This confirms a clear and significant advantage over Method \Rmnum{2}. When the three most important genes were used, Method \Rmnum{1} achieved a 118$\%$-186$\%$ increase in successful cases compared to Method \Rmnum{2}. It also matched up to 95$\%$ of success cases of Method \Rmnum{2}, while continuing to reduce computation time, highlighting both robustness and efficiency. Among the successful cases obtained only by Method \Rmnum{2}, the average runtime is greater than the average runtime of all successful cases by Method \Rmnum{2}. This suggests the potential complexity of these cases, which may explain why Method \Rmnum{1} failed to produce results within the time limit. Furthermore, comparison across different gene set sizes shows that selecting more important genes steadily improved performance in both the (1) and (2) categories. Notably, the results of cases in (1) showed a more than threefold increase, demonstrating the framework’s effectiveness in identifying additional, previously undetected successful cases. These findings highlight the scalability, efficiency, and effectiveness of the proposed framework. 

\begin{table*}[hbtp]
\captionsetup{justification=centering}
\caption{Performance comparison on iML1515 dataset}
\label{t2}
\centering
\scriptsize
\begin{threeparttable}
\begin{tabular}{cccccccc}
\toprule
 \multicolumn{2}{c}{Dataset iML1515} & \multicolumn{2}{c}{(1)Only Method \Rmnum{1} Succeeded} & \multicolumn{2}{c}{(2)Both Methods Succeeded} & \multicolumn{2}{c}{(3)Only Method \Rmnum{2} Succeeded} \\
\cmidrule(lr){1-2} \cmidrule(lr){3-4} \cmidrule(lr){5-6} \cmidrule(lr){7-8}
Size \tnote{a} & Method \Rmnum{1} \tnote{b} & Succ. Case & Avg. Time & Succ. Case & Avg. Time & Succ. Case & Avg. Time\\
\midrule
1 & St1 to St5 & 62 & 354.71 & 149 & 271.97 & 19 & 246.71 \\
\midrule
2 & St1 to St5 & 81 & 351.31 & 147 & 287.27 & 21 & 306.37 \\
\midrule
3 & St1 to St5 & 81 & 326.22 & 147 & 292.92 & 21 & 231.15 \\
\midrule
\bottomrule
& Method \Rmnum{2}  & 168 & 283.76  & 168 & 283.76 & 168 & 283.76 \\
\midrule
\end{tabular}
\begin{tablenotes}
\tiny
\item[a] The number of important genes selected for Method \Rmnum{1}.
\item[b] The time limit for each metabolite is 500 seconds.
\end{tablenotes}
\end{threeparttable}
\end{table*}

\begin{table}[htbp]
\captionsetup{justification=centering}
\caption{Performance Comparison on e$\_$coli$\_$core Dataset}
\label{t3}
\centering
\footnotesize
\begin{threeparttable}
\begin{tabular}{ccccccc}
\toprule
 \multicolumn{2}{c}{Dataset e$\_$coli$\_$core} & \multicolumn{2}{c}{Only Method \Rmnum{1} Succeeded} & \multicolumn{3}{c}{Both Method \Rmnum{1} and Method \Rmnum{2} Succeeded} \\
\cmidrule(lr){1-2} \cmidrule(lr){3-4} \cmidrule(lr){5-7}
Size \tnote{a} & Method \Rmnum{1} \tnote{b} & Succ. Case & Avg. Time & Succ. Case & Ratio & Avg. Time \\
\midrule
\multirow{3}{*}{$1$}
& St1,St2,St3 & 0 & - & 45 & 97.83$\%$ & 236.89 \\
& St4 & 0 & - & 44 & 95.65$\%$ & 288.01 \\
& St5 & 2 & 365.49 & 44 & 95.65$\%$ & 329.37 \\
\midrule
\multirow{3}{*}{$2$}
& St1,St2,St3 & 0 & - & 45 & 97.83$\%$ & 250.35 \\
& St4 & 0 & - & 45 & 97.83$\%$ & 366.25 \\
& St5 & 0 & - & 45 & 97.83$\%$ & 347.03 \\
\midrule
\multirow{2}{*}{$3$}
& St1,St2,St3,St4 & 1 & 187.10 & 45 & 97.83$\%$ & 364.95 \\
& St5 & 0 & - & 45 & 97.83$\%$ & 371.67 \\
\midrule
\midrule
\multicolumn{2}{c}{Method \Rmnum{2}}  & 46 & 266.87  & 46 & - & 266.87 \\
\bottomrule
\end{tabular}
\begin{tablenotes}
\tiny
\item[a] The number of important genes selected for Method \Rmnum{1}.
\item[b] The time limit for each metabolite is 500 seconds.
\end{tablenotes}
\end{threeparttable}
\end{table}

Table~\ref{t2} presents the performance comparison between Method \Rmnum{1} and Method \Rmnum{2}, each using three different sizes of ranked gene sets, on the iML1515 dataset under a specified time limit. Notably, the top three genes were ranked identically by all five strategies for this dataset. When one important gene was included, Method \Rmnum{1} outperformed Method \Rmnum{2} by producing 36.90$\%$ additional successful cases, indicating a notable improvement. It also preserved 88.69$\%$ of successful cases of Method \Rmnum{2} while reducing runtime, demonstrating a clear advantage in both effectiveness and efficiency. With two important genes, Method \Rmnum{1} achieved a 48.21$\%$ increase in successful cases compared to Method \Rmnum{2} and reliably reproduced nearly 90$\%$ of Method \Rmnum{2}’s results, further confirming its robustness and superior performance. Using three important genes, Method \Rmnum{1} generated almost 50$\%$ more successful cases than Method \Rmnum{2}. It also maintained over 87.50$\%$ overlap with Method \Rmnum{2}’s success cases while achieving a comparable average runtime, highlighting both its computational efficiency and reliability. 

\begin{figure}[htbp]
    \centering
    \includegraphics[width=0.6\textwidth, trim=20pt 20pt 10pt 10pt, clip]{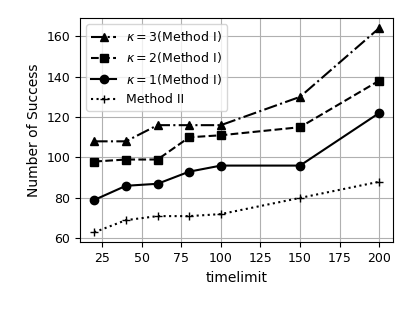}
    \caption{Number of successful outcomes under different time limits for Method \Rmnum{1} and Method \Rmnum{2}.}
    \label{fig2}
\end{figure}

The results on the e$\_$coli$\_$core dataset, a small-scale constraint-based metabolic network comprising only 72 metabolites and 95 reactions, are provided in Table~\ref{t3}. On this dataset, Method \Rmnum{1} yielded a few additional success cases. However, they were still able to recover nearly all the success cases identified by Method \Rmnum{2}. This outcome suggests that, for small-scale models, the primary bottleneck is not computation time, and Method \Rmnum{2} may already operate at full efficiency. Furthermore, the observed increase in average runtime indicates that the proposed framework does not improve computational efficiency in small-scale models. For the few successful cases obtained solely through Method \Rmnum{2}, their average runtime largely exceeded the mean runtime of all Method \Rmnum{2}'s successful cases. 

\begin{table}[htbp]
\centering
\begin{threeparttable}
\caption{Comparison of success rates on iMM904 dataset with OptGene}
\label{tab:optgene}
\begin{tabular}{ccc}
\toprule
Top-k gene \tnote{a} & Strategy \tnote{b} & iMM904 \\
\midrule
\multirow{1}{*}{$1$}
& St1, St2, St3, St4, St5 & 7.16\% \\
\midrule
\multirow{2}{*}{$2$}
& St1, St2 & 7.54\% \\
& St3, St4, St5 & 7.80\% \\
\midrule
\multirow{3}{*}{$3$}
& St1, St2 & 8.18\% \\
& St3, St4 & 8.31\% \\
& St5  & 7.80\% \\
\midrule
& Original OptGene \tnote{b} & 6.65\% \\
\bottomrule
\end{tabular}
\begin{tablenotes}
\tiny
\item[a] Selected Top-k important genes associating with Gene-Ranked OptGene.
\item[b] Maximum global time limit for each metabolite is 10 seconds.
\end{tablenotes}
\end{threeparttable}
\end{table}

\begin{table}[htbp]
\centering
\begin{threeparttable}
\caption{Comparison of success rates across different models with GDLS}
\label{tab:gdls}
\begin{tabular}{lccc}
\toprule
Strategy & e\_coli\_core & iAF1260 & iJO1366 \\
\midrule
St1 to St5 \tnote{a,b} & 5.77\% & 0.42\% & 0.38\% \\
Original GDLS \tnote{b}         & 5.77\% & 0.42\% & 0.29\% \\
\bottomrule
\end{tabular}
\begin{tablenotes}
\tiny
\item[a] Selected three important genes associated with Gene-Ranked GDLS.
\item[b] Maximum time limit for each metabolite is 100 seconds.
\end{tablenotes}
\end{threeparttable}
\end{table}

Figure~\ref{fig2} illustrates the number of successful outcomes achieved under different time limits for three Method \Rmnum{1} variants and Method \Rmnum{2} on the iML1515 dataset. Overall, the number of successes increased as the time limit extended for all strategies, indicating a positive correlation between computational budget and performance. Among the compared approaches, Method \Rmnum{1} with three important genes consistently achieved the highest number of successes across all time limits, demonstrating superior overall performance. Its advantage became more pronounced at longer time limits, particularly at 200 seconds, where a substantial improvement over the other methods is observed.  Method \Rmnum{1} with two important genes ranked second and exhibited steady growth as the time limit increased.  Method \Rmnum{1} with one important gene also showed an upward trend, but its performance remained consistently lower than that of the variants of two and three genes. In contrast, Method \Rmnum{2} yielded the lowest performance across all settings, although it similarly benefited from increased time limits. These results suggest that the proposed gene ranking strategies more effectively leverage additional computational resources, leading to significantly improved success rates compared with the  Method \Rmnum{2}.

\begin{table}[htbp]
\centering
\caption{Four Straindesign tasks with GLPK on e$\_$coli$\_$core model}
\label{gmcs:1}
\begin{threeparttable}
\begin{tabular}{ccccc}
\toprule
Strategy & SUCP & dGCP & wGCP & pGCP \\
\midrule
St1, St2, St3 & 66.67\% & 50.00\% & 66.67\% & 100\% \\
St4  & 66.67\% & 50.00\% & 58.33\% & 100\% \\
St5 & 75.00 & 58.33\% & 66.67\% & 100\% \\
\midrule
$\emptyset$ & 58.33\% & 41.67\% & 58.33\% & 100\% \\
\bottomrule
\end{tabular}
\end{threeparttable}
\end{table}

\begin{table}[htbp]
\centering
\caption{Four Straindesign tasks with SCIP on e$\_$coli$\_$core model}
\label{gmcs:2}
\begin{threeparttable}
\begin{tabular}{ccccc}
\toprule
Strategy & SUCP & dGCP & wGCP & pGCP \\
\midrule
St1, St2, St3 & 66.67\% & 58.33\% &58.33\% & 100\% \\
St4  & 66.67\% & 58.33\% & 66.67\% & 100\% \\
St5 & 75.00 & 58.33\% & 66.67\% & 100\% \\
\midrule
$\emptyset$ & 66.67\% & 50.00\% & 58.33\% & 100\% \\
\bottomrule
\end{tabular}
\end{threeparttable}
\end{table}

\begin{table}[htbp]
\centering
\caption{Four straindesign tasks}
\label{table:task}
\begin{threeparttable}
\begin{tabular}{cc}
\toprule
 Abbr. & Task \\
\midrule
SUCP & substrate-uptake-coupled production. \\
dGCP &  directionally growth-coupled production. \\
wGCP & weakly growth-coupled production. \\
pGCP & potentially growth-coupled production. \\
\bottomrule
\end{tabular}
\end{threeparttable}
\end{table}

\subsection{Gene ranking strategies for other methods}
To evaluate the effectiveness of the gene ranking framework as a plug-and-play enhancement for other constraint-based model design methods, computational experiments were conducted using OptGene and GDLS \cite{lun2009large, patil2005evolutionary}, as well as the gMCS-based methods provided by StrainDesign \cite{schneider2022straindesign}. Because the baseline success rates of OptGene and GDLS were low, the improvement achieved by the gene ranking framework was limited. Results are provided in Tables~\ref{tab:optgene} and ~\ref{tab:gdls}.

\begin{table*}[htbp]
\captionsetup{justification=centering}
\caption{Solving Process of MILPs for the Target Metabolites where Both Methods Succeeded}
\label{t4}
\centering
\scriptsize
\begin{threeparttable}
\begin{tabular}{cccccccc}
\toprule
\multicolumn{2}{c}{Dataset iMM904} & \multicolumn{6}{c}{Key Statistics of Solving Process (Cases where Both Methods were successful)} \\
\cmidrule(lr){1-2} \cmidrule(lr){3-8}
Size\tnote{a} & Method \Rmnum{1} & Avg. Nodes\tnote{b} & Avg. Rows\tnote{c} & Avg. Columns\tnote{c} & Avg. Non-zeros\tnote{d} & Avg. Binary\tnote{d} & Avg. Cuts\tnote{e} \\
\midrule
\multirow{2}{*}{$1$}
& St1,St2,St4 & 1363.91 & 1141.04 & 1037.25 & 4593.85 & 448.85 & 19.02 \\
& St3,St5 & 1477.21 & 1171.94 & 1062.24 & 4750.59 & 447.71 & 18.32 \\
\midrule
\multirow{4}{*}{$2$}
& St1,St2 & 647.03 & 1145.42 & 1040.50 & 4628.47 & 450.87 & 14.32 \\
& St3 & 1439.94 & 1158.94 & 1053.74 & 4718.00 & 439.87 & 12.61 \\
& St4 & 3412.75 & 1119.94 & 1023.24 & 4478.72 & 441.42 & 42.03\\
& St5 & 3143.09 & 1109.16 & 1019.81 & 4450.63 & 432.74 & 26.37 \\
\midrule
\multirow{3}{*}{$3$}
& St1,St2 & 3543.06 & 1138.86 & 1037.22 & 4599.61 & 448.81 & 16.61 \\
& St3,St4 & 1476.85 & 1117.49 & 1023.95 & 4502.34 & 436.34 & 17.73 \\
& St5 & 1646.30 & 1120.87 & 1030.34 & 4511.09 & 432.34 & 12.70 \\
\midrule
\bottomrule
& Method \Rmnum{2}  & 5006.71 & 1162.71  & 1046.49 & 4613.01 & 447.90 & 56.36 \\
\midrule
\end{tabular}
\begin{tablenotes}
\tiny
\item[a] The number of important genes selected for Method \Rmnum{1}.
\item[b] The average number of nodes searched in branch-and-bound trees, and the smaller the better.
\item[c] The average number of constraints and variables after pro-solving processes, and the smaller the better. Rows correspond to constraints, and columns correspond to variables in a constraint matrix.
\item[d] The average number of non-zero counts and binary variables, and the smaller the better.
\item[e] The average number of cutting planes, and the smaller the better.
\end{tablenotes}
\end{threeparttable}
\end{table*}

The gMCS-based method was applied for four StrainDesign tasks for growth-coupled production (SUCP, dGCP, wGCP, and pGCP) to the e$\_$coli$\_$core model for target metabolites with excretion reactions. Definitions of the four StrainDesign tasks are listed in Table~\ref{table:task}. Although these task definitions differ from those in this study, the experiments are useful for assessing the effectiveness of the gene ranking framework. Tables \ref{gmcs:1} and \ref{gmcs:2} present results obtained with and without the gene ranking framework using two MILP solvers, GLPK and SCIP \cite{bolusani2024scip}. The results show a consistent 100\% success rate for the pGCP task across all strategies and the baseline on both solvers. Strategy St5 improved performance in the other three tasks compared with the baseline on both solvers, reaching 75.00\% in the SUCP task, which is higher than the original gMCS-based method (GLPK: 58.33\%, SCIP: 66.67\%). In addition, strategies St1–St4 outperformed the baseline in the dGCP task on both solvers, suggesting improved capability for these tasks.

\section{Discussion and Conclusion}\label{sec:discuss}
\subsection{Gene Ranking Framework}
As previously described, identifying gene deletion strategies for growth-coupled production within limited computational time remains a major challenge, particularly in genome-scale constraint-based metabolic networks, where success rates are often low. In this study, we developed a method to rank genes based on GPR associations and network topological features. By fixing these highly ranked genes to binary values (0 or 1) at an early stage and solving the resulting MILP subproblems in parallel, we improved the success rate of identifying feasible strategies within a given time limit. Since MILP solvers already employ internal parallelization even without this framework, the observed improvement demonstrates the effectiveness of the prioritization scheme introduced by the gene ranking framework.

In the computational experiments, we compared the performance of the gene ranking framework when applied (Method \Rmnum{1}) and not applied (Method \Rmnum{2}). As the baseline for Method II, we used RatGene, a representative MILP-based method for computing gene deletion strategies for growth-coupled production. The results showed that, in genome-scale constraint-based metabolic models, there were many cases in which only Method I was able to identify feasible gene deletion strategies within the given time limit. These results demonstrate the effectiveness of the proposed approach.

\subsection{Investigation of the MILP Solving Process}
Computational results in Tables~\ref{t1} and \ref{t2} show that Method \Rmnum{1} consistently achieved at least 80$\%$ of the successful cases found by Method \Rmnum{2}. 
Furthermore, Method \Rmnum{1} increased the number of successful cases within the same time limit by 48\% for iML1515 and 186\% for iMM904 compared to Method \Rmnum{2}. These improvements can be attributed to the assignment of binary values to important gene variables, which effectively altered the structure of the MILP problems. 

To further analyze the impact of Method \Rmnum{1} on the optimization process, we examined how the MILP problems reformulated by Method \Rmnum{1} differed from those formulated by Method \Rmnum{2} during the solution process, using the iMM904 dataset as a case study. Modern MILP solvers employ several powerful techniques to efficiently solve problems, including presolving, branch and bound, and cutting planes. Presolving simplifies the problem by eliminating duplicate rows, removing redundant columns, and fixing the values of certain variables prior to solving. Solvers also store and process only non-zero coefficients to optimize memory and computation \cite{bixby1999mip}. The branch-and-bound technique iteratively constructs a search tree, solving LP relaxations at each node, and applies bounding and pruning strategies to reduce the search space \cite{land2009automatic}. The cutting plane method improves the formulation by progressively adding valid constraints that tighten the feasible region, thereby guiding the solver more efficiently toward the optimal solution \cite{kelley1960cutting}.

\begin{table*}[htbp]
\captionsetup{justification=centering}
\caption{Solving Process of MILP Problems from Success Cases Where Only Method \Rmnum{2} Succeeded}
\label{t5}
\centering
\scriptsize
\begin{threeparttable}
\begin{tabular}{cccccccc}
\toprule
\multicolumn{2}{c}{Dataset iMM904} & \multicolumn{6}{c}{Key Statistics of Solving Process (Cases where only Method \Rmnum{2} were successful)} \\
\cmidrule(lr){1-2} \cmidrule(lr){3-8}
Size\tnote{a} & Method \Rmnum{1} & Avg. Nodes\tnote{b} & Avg. Rows\tnote{c} & Avg. Columns\tnote{c} & Avg. Non-zeros\tnote{d} & Avg. Binary\tnote{d} & Avg. Cuts\tnote{e} \\
\midrule
\multirow{2}{*}{$1$}
& St1,St2,St4 & 7564.88 & 1117.76 & 1017.68 & 4390.54 & 435.73 & 60.54 \\
& St3,St5 & 6482.92 & 1115.34 & 1020.15 & 4416.29 & 423.25 & 45.85 \\
\midrule
\multirow{4}{*}{$2$}
& St1,St2 & 7145.48 & 1102.37 & 1007.38 & 4336.15 & 433.26 & 61.98 \\
& St3 & 6972.36 & 1113.88 & 1015.23 & 4367.59 & 422.08 & 55.22 \\
& St4 & 7585.82 & 1089.95 & 997.19 & 4277.57 & 429.46 & 55.29\\
& St5 & 7706.75 & 1070.42 & 990.98 & 4203.45 & 415.71 & 63.32 \\
\midrule
\multirow{3}{*}{$3$}
& St1,St2 & 6647.53 & 1105.34 & 1010.02 & 4349.61 & 433.43 & 47.90 \\
& St3,St4 & 6685.82 & 1095.92 & 1005.43 & 4307.69 & 423.82 & 50.40 \\
& St5 & 6392.79 & 1091.18 & 1005.17 & 4286.86 & 421.46 & 45.61 \\
\midrule
\bottomrule
& Method \Rmnum{2}  & 5006.71 & 1162.71  & 1046.49 & 4613.01 & 447.90 & 56.36 \\
\midrule
\end{tabular}
\begin{tablenotes}
\tiny
\item[a] The number of important genes selected for Method \Rmnum{1}.
\item[b] The average number of nodes searched in branch-and-bound trees, and the smaller the better.
\item[c] The average number of constraints and variables after pro-solving processes, and the smaller the better. Rows correspond to constraints, and columns correspond to variables in a constraint matrix.
\item[d] The average number of non-zero counts and binary variables, and the smaller the better.
\item[e] The average number of cutting planes, and the smaller the better.
\end{tablenotes}
\end{threeparttable}
\end{table*}
Table~\ref{t4} presents key statistics of the MILP solution process constructed by Method \Rmnum{1} and \Rmnum{2}, using cases where both methods were successful on the iMM904 dataset. Across all configurations assigning one, two, or three important genes, Method \Rmnum{1} significantly explored fewer branch-and-bound nodes than Method \Rmnum{2}, as indicated by the Avg. Nodes metric in Table~\ref{t4}. This reduction in LP relaxations was a major factor contributing to improved computational efficiency. Method \Rmnum{1} also slightly reduced the problem size in many cases, as evidenced by decreases in Avg. Rows (constraints), Avg. Columns (variables), and Avg. Non-zeros reported in Table~\ref{t4}. In particular, when three important genes were assigned, all these metrics were lower than the averages of Method \Rmnum{2} and also outperformed the configurations with one or two genes. This suggests that fixing binary values for a larger number of key gene variables simplifies the problem structure and facilitates the solution. Furthermore, Method \Rmnum{1} generated significantly fewer cutting planes than Method \Rmnum{2}, as reflected by the reduced Avg. Cuts metric in Table~\ref{t4}.
These results support the conclusion that Method \Rmnum{1} reduces MILP problem complexity and enables faster solutions within the time limit. These findings are consistent with the performance results reported for set (2) in Table~\ref{t1} and provide insight into the sources of improvement. However, the relationship between MILP complexity and runtime is not strictly linear. For example, in the \hyperref[e:m]{St1} and \hyperref[e:f]{St2} strategies using two important genes, the average number of explored nodes was substantially lower than that of Method \Rmnum{2}. Nevertheless, due to the internal behavior of MILP solvers, including heuristic strategies, secondary branch-and-bound processes, and other internal procedures, the reduction in nodes did not always result in a proportional decrease in average runtime, as shown in Table~\ref{t1}. More detailed statistics are provided in the supplementary Table S1.

Method \Rmnum{1} essentially partitioned the problem constructed by original RatGene (Method \Rmnum{2}) into mutually exclusive subproblems for parallel computation, and then solved each subproblem independently. In principle, the successful results obtained by Method \Rmnum{2} should be fully reproduced by combining the solutions of all subproblems by Method \Rmnum{1}. However, due to the computation time limit, the ratios reported in Table \ref{t1} did not reach 100$\%$. The following discussion examines the causes of these non-reproducible cases. 

Table~\ref{t5} presents key statistics on the MILP solution process by Methods \Rmnum{1} and \Rmnum{2}, focusing on the cases where knockout strategies were identified by Method \Rmnum{2} but not by Method \Rmnum{1}. In terms of the average number of explored nodes, Method \Rmnum{1} required more extensive search than Method \Rmnum{2}, suggesting that fixing certain gene variables made the problems more difficult to solve. However, Method \Rmnum{1} consistently produced smaller MILP instances than Method \Rmnum{2} across four metrics: average number of rows (constraints), columns (variables), non-zero entries, and binary variables. In addition, when three important genes were fixed, Method \Rmnum{1} also resulted in fewer cutting planes than Method \Rmnum{2}. These results indicate that the main reason why some deletion strategies could not be reproduced by Method \Rmnum{1} within the time limit is the increased complexity of the branch-and-bound trees. Furthermore, comparisons across different numbers of fixed important genes show that all complexity measures improved as more gene variables were fixed. This is consistent with Table \ref{t4} and supports that fixing more gene variables makes the problems easier to solve. More detailed statistics are provided in the supplementary Table S2.

Thus, although there is a correlation between the number of gene variables and the complexity of the constructed problems, there is no explicit linear relationship between them. Moreover, because modern MILP solvers operate through workflows that integrate multiple internal mechanisms, a reduction in the number of search nodes does not necessarily lead to a proportional decrease in average runtime. In addition, even when two genes are assigned equal importance by the proposed framework, independently fixing them to the same value does not necessarily result in identical MILP problem structures. This is because the proposed method does not fully incorporate all information from the network. Furthermore, although extremely rare, even when the constructed problems are identical, the reduction in search nodes achieved by assigning values to both genes is not simply twice that obtained by assigning a single gene. This is because problem construction also depends on other aspects of the network.

\begin{table}[htbp]
\caption{Solving Process of MILP Problems by HiGHS on iMM904 dataset for the Target Metabolites where Both Methods Succeeded}
\label{t:highs}
\centering
\scriptsize
\begin{threeparttable}
\begin{tabular}{cccccccccc}
\toprule
Method \Rmnum{1} & Rows & Columns & Nonzeros & Continuous \\
\midrule
St1, St2, St4 & 1264.37  & 1168.92  & 4860.01 & 593.94  \\
St3, St5 & 1254.88  & 1157.88  & 4811.67 & 587.41  \\
\midrule
Method \Rmnum{2} & 1266.39  & 1169.88  & 4862.32 & 594.48   \\
\midrule
\midrule
Method \Rmnum{1} & Binaries & Explored nodes & Simplex Iterations & Runtime \\
\midrule
St1, St2, St4 & 549.58  & 2081.42  & 402248.87  & 275.38 \\
St3, St5 & 545.47  & 2001.24 & 398834.51   & 220.20  \\
\midrule
Method \Rmnum{2} & 550.04  & 2088.98  & 393867.21 & 315.84  \\
\bottomrule
\end{tabular}
\end{threeparttable}
\end{table}

We also evaluated the MILP-solving performance when applying the gene-ranking framework using different solvers. Tables \ref{t:highs} and \ref{t:gurobi} present detailed comparisons of MILP statistics for the target compounds in iMM904 for which Method \Rmnum{1} and Method \Rmnum{2} successfully identified valid gene deletion strategies (key statistics are provided in supplementary Tables S3 and S4).
The evaluation metrics include the number of constraints (rows), the number of variables (columns), the number of nonzero entries (nonzeros), the number of continuous variables (continuous), the number of binary variables (binaries), the number of integer variables (integers), the number of branch-and-bound nodes (explored nodes), the number of simplex iterations (simplex iterations), the number of cutting planes (cuts), and the average computational time (runtime). Results from HiGHS \cite{huangfu2018parallelizing} showed that, although the overall improvements in the metrics were relatively modest, the computational time was substantially reduced. Results from Gurobi \cite{gurobi} showed that applying the gene-ranking framework substantially reduced the size of the search tree (explored nodes) and decreased the number of simplex iterations. Among the three solvers, Gurobi achieved the fastest runtime; however, the reduction in computational time obtained by applying the gene-ranking framework was relatively limited. This may suggest that the acceleration strategies implemented in Gurobi for general MILP problems are consistent with the acceleration approach introduced by the gene-ranking framework.

\begin{table*}[htbp]
\caption{Solving Process of MILP Problems by Gurobi on iMM904 dataset for the Target Metabolites where Both Methods Succeeded}
\label{t:gurobi}
\centering
\scriptsize
\begin{threeparttable}
\begin{tabular}{cccccccccc}
\toprule
Method \Rmnum{1} & Rows & Columns & Nonzeros & Binaries & Integers & Explored nodes & Simplex Iterations & Cuts  & Runtime \\
\midrule
St1, St2, St4 & 930.56 & 919.62 & 4014.53 & 414.00 & 414.00 & 184738.41 & 6555716.10 & 580.91 & 107.00 \\
St3, St5 & 922.26 & 910.63 & 3977.84 & 410.68 & 410.68 & 123037.01 & 4438226.38 & 585.95 & 102.44 \\
\midrule
Method \Rmnum{2} & 931.31 & 920.57 & 4018.51 & 414.29 & 414.29  & 186595.88 & 6576230.90 & 570.60 & 107.83 \\
\bottomrule
\end{tabular}
\end{threeparttable}
\end{table*}

\subsection{Conclusion}
In this study, we propose a gene ranking framework for metabolic design using constraint-based models, which reduces the complexity of MILP problems by prioritizing and fixing values of important gene variables. To this end, we introduce five strategies to score and rank gene importance by leveraging GPR associations and network topology. The proposed method discovered many successful cases that could not be solved by the baseline method within the same time limit. Although problem complexity increased in some cases, it was consistently reduced overall as more gene variable values were fixed. Furthermore, the proposed gene ranking strategies can be seamlessly integrated into other constraint-based methods, in a plug-and-play manner, leading to improved efficiency under identical time constraints. When combined with gMCS-based methods, the framework also improved performance in several growth-coupled production tasks compared to the standard gMCS-based approach. Overall, this study demonstrates that leveraging prior information to assign gene variable values can substantially reduce problem complexity and accelerate solution processes without major modifications to existing algorithms. These findings suggest strong potential for application to other time-sensitive constraint-based methods.

\bibliographystyle{unsrt}
\bibliography{document}

@article{schneider2022straindesign,
  title={StrainDesign: a comprehensive Python package for computational design of metabolic networks},
  author={Schneider, Philipp and Bekiaris, Pavlos Stephanos and von Kamp, Axel and Klamt, Steffen},
  journal={Bioinformatics},
  volume={38},
  number={21},
  pages={4981--4983},
  year={2022},
  publisher={Oxford University Press}
}

@article{ranganathan2010optforce,
  title={OptForce: an optimization procedure for identifying all genetic manipulations leading to targeted overproductions},
  author={Ranganathan, Sridhar and Suthers, Patrick F and Maranas, Costas D},
  journal={PLoS Comput Biol},
  volume={6},
  number={4},
  pages={e1000744},
  year={2010},
  publisher={Public Library of Science}
}

@article{orth2010reconstruction,
  title={Reconstruction and use of microbial metabolic networks: the core {\it Escherichia coli} metabolic model as an educational guide},
  author={Orth, Jeffrey D and Fleming, Ronan MT and Palsson, Bernhard O},
  journal={EcoSal plus},
  year={2010},
  publisher={American Society for Microbiology}
}

@book{maranas2016optimization,
  title={Optimization methods in metabolic networks},
  author={Maranas, Costas D and Zomorrodi, Ali R},
  year={2016},
  publisher={John Wiley \& Sons}
}

@article{orth2011comprehensive,
  title={A comprehensive genome-scale reconstruction of {\it Escherichia coli} metabolism—2011},
  author={Orth, Jeffrey D and Conrad, Tom M and Na, Jessica and Lerman, Joshua A and Nam, Hojung and Feist, Adam M and Palsson, Bernhard {\O}},
  journal={Molecular systems biology},
  volume={7},
  number={1},
  pages={535},
  year={2011},
  publisher={John Wiley \& Sons, Ltd Chichester, UK}
}

@article{heirendt2019creation,
  title={Creation and analysis of biochemical constraint-based models using the COBRA Toolbox v. 3.0},
  author={Heirendt, Laurent and Arreckx, Sylvain and Pfau, Thomas and Mendoza, Sebasti{\'a}n N and Richelle, Anne and Heinken, Almut and Haraldsd{\'o}ttir, Hulda S and Wachowiak, Jacek and Keating, Sarah M and Vlasov, Vanja and others},
  journal={Nature protocols},
  volume={14},
  number={3},
  pages={639--702},
  year={2019},
  publisher={Nature Publishing Group}
}

@article{orth2010flux,
  title={What is flux balance analysis?},
  author={Orth, Jeffrey D and Thiele, Ines and Palsson, Bernhard {\O}},
  journal={Nature biotechnology},
  volume={28},
  number={3},
  pages={245--248},
  year={2010},
  publisher={Nature Publishing Group}
}

@article{feist2007genome,
  title={A genome-scale metabolic reconstruction for {\it Escherichia coli} K-12 MG1655 that accounts for 1260 ORFs and thermodynamic information},
  author={Feist, Adam M and Henry, Christopher S and Reed, Jennifer L and Krummenacker, Markus and Joyce, Andrew R and Karp, Peter D and Broadbelt, Linda J and Hatzimanikatis, Vassily and Palsson, Bernhard {\O}},
  journal={Molecular systems biology},
  volume={3},
  number={1},
  pages={121},
  year={2007},
  publisher={EMBO Press}
}

@article{burgard2003optknock,
  title={Optknock: a bilevel programming framework for identifying gene knockout strategies for microbial strain optimization},
  author={Burgard, Anthony P and Pharkya, Priti and Maranas, Costas D},
  journal={Biotechnology and bioengineering},
  volume={84},
  number={6},
  pages={647--657},
  year={2003},
  publisher={Wiley Online Library}
}

@article{patil2005evolutionary,
  title={Evolutionary programming as a platform for in silico metabolic engineering},
  author={Patil, Kiran Raosaheb and Rocha, Isabel and F{\"o}rster, Jochen and Nielsen, Jens},
  journal={BMC bioinformatics},
  volume={6},
  number={1},
  pages={308},
  year={2005},
  publisher={BioMed Central}
}

@article{lun2009large,
  title={Large-scale identification of genetic design strategies using local search},
  author={Lun, Desmond S and Rockwell, Graham and Guido, Nicholas J and Baym, Michael and Kelner, Jonathan A and Berger, Bonnie and Galagan, James E and Church, George M},
  journal={molecular systems biology},
  volume={5},
  number={1},
  pages={296},
  year={2009},
  publisher={EMBO Press}
}

@article{schneider2021systematizing,
  title={Systematizing the different notions of growth-coupled product synthesis and a single framework for computing corresponding strain designs},
  author={Schneider, Philipp and Mahadevan, Radhakrishnan and Klamt, Steffen},
  journal={Biotechnology Journal},
  volume={16},
  number={12},
  pages={2100236},
  year={2021},
  publisher={Wiley Online Library}
}

@article{tamura2023gene,
  title={Gene deletion algorithms for minimum reaction network design by mixed-integer linear programming for metabolite production in constraint-based models: gDel\_minRN},
  author={Tamura, Takeyuki and Muto-Fujita, Ai and Tohsato, Yukako and Kosaka, Tomoyuki},
  journal={Journal of Computational Biology},
  year={2023},
  publisher={Mary Ann Liebert, Inc., publishers 140 Huguenot Street, 3rd Floor New~…}
}

@article{norsigian2020bigg,
  title={BiGG Models 2020: multi-strain genome-scale models and expansion across the phylogenetic tree},
  author={Norsigian, Charles J and Pusarla, Neha and McConn, John Luke and Yurkovich, James T and Dr{\"a}ger, Andreas and Palsson, Bernhard O and King, Zachary},
  journal={Nucleic acids research},
  volume={48},
  number={D1},
  pages={D402--D406},
  year={2020},
  publisher={Oxford University Press}
}

@article{yasemi2021modelling,
  title={Modelling cell metabolism: a review on constraint-based steady-state and kinetic approaches},
  author={Yasemi, Mohammadreza and Jolicoeur, Mario},
  journal={Processes},
  volume={9},
  number={2},
  pages={322},
  year={2021},
  publisher={MDPI}
}

@article{nielsen2017systems,
  title={Systems biology of metabolism},
  author={Nielsen, Jens},
  journal={Annual review of biochemistry},
  volume={86},
  number={1},
  pages={245--275},
  year={2017},
  publisher={Annual Reviews}
}

@article{kadir2010modeling,
  title={Modeling and simulation of the main metabolism in Escherichia coli and its several single-gene knockout mutants with experimental verification},
  author={Kadir, Tuty Asmawaty Abdul and Mannan, Ahmad A and Kierzek, Andrzej M and McFadden, Johnjoe and Shimizu, Kazuyuki},
  journal={Microbial cell factories},
  volume={9},
  number={1},
  pages={88},
  year={2010},
  publisher={Springer}
}

@article{khodayari2014kinetic,
  title={A kinetic model of Escherichia coli core metabolism satisfying multiple sets of mutant flux data},
  author={Khodayari, Ali and Zomorrodi, Ali R and Liao, James C and Maranas, Costas D},
  journal={Metabolic engineering},
  volume={25},
  pages={50--62},
  year={2014},
  publisher={Elsevier}
}

@article{kim2012recent,
  title={Recent advances in reconstruction and applications of genome-scale metabolic models},
  author={Kim, Tae Yong and Sohn, Seung Bum and Kim, Yu Bin and Kim, Won Jun and Lee, Sang Yup},
  journal={Current opinion in biotechnology},
  volume={23},
  number={4},
  pages={617--623},
  year={2012},
  publisher={Elsevier}
}

@book{stephanopoulos1998metabolic,
  title={Metabolic engineering: principles and methodologies},
  author={Stephanopoulos, George and Aristidou, Aristos A and Nielsen, Jens},
  year={1998},
  publisher={Elsevier}
}

@article{schauer1983quasi,
  title={Quasi-steady-state approximation in the mathematical modeling of biochemical reaction networks},
  author={Schauer, M and Heinrich, R},
  journal={Mathematical biosciences},
  volume={65},
  number={2},
  pages={155--170},
  year={1983},
  publisher={Elsevier}
}

@article{gombert2000mathematical,
  title={Mathematical modelling of metabolism},
  author={Gombert, Andreas Karoly and Nielsen, Jens},
  journal={Current opinion in biotechnology},
  volume={11},
  number={2},
  pages={180--186},
  year={2000},
  publisher={Elsevier}
}

@article{gottstein2016constraint,
  title={Constraint-based stoichiometric modelling from single organisms to microbial communities},
  author={Gottstein, Willi and Olivier, Brett G and Bruggeman, Frank J and Teusink, Bas},
  journal={Journal of the Royal Society Interface},
  volume={13},
  number={124},
  pages={20160627},
  year={2016},
  publisher={The Royal Society}
}

@article{ma2025ratgene,
  title={RatGene: Gene deletion-addition algorithms using growth to production ratio for growth-coupled production in constraint-based metabolic networks},
  author={Ma, Yier and Tamura, Takeyuki},
  journal={IEEE Transactions on Computational Biology and Bioinformatics},
  year={2025},
  publisher={IEEE}
}

@incollection{land2009automatic,
  title={An automatic method for solving discrete programming problems},
  author={Land, Ailsa H and Doig, Alison G},
  booktitle={50 Years of Integer Programming 1958-2008: From the Early Years to the State-of-the-Art},
  pages={105--132},
  year={2009},
  publisher={Springer}
}

@article{kelley1960cutting,
  title={The cutting-plane method for solving convex programs},
  author={Kelley, Jr, James E},
  journal={Journal of the society for Industrial and Applied Mathematics},
  volume={8},
  number={4},
  pages={703--712},
  year={1960},
  publisher={SIAM}
}

@inproceedings{bixby1999mip,
  title={MIP: Theory and practice—closing the gap},
  author={Bixby, E Robert and Fenelon, Mary and Gu, Zonghao and Rothberg, Ed and Wunderling, Roland},
  booktitle={IFIP Conference on System Modeling and Optimization},
  pages={19--49},
  year={1999},
  organization={Springer}
}

@article{fleming2012mass,
  title={Mass conserved elementary kinetics is sufficient for the existence of a non-equilibrium steady state concentration},
  author={Fleming, Ronan MT and Thiele, Ines},
  journal={Journal of Theoretical Biology},
  volume={314},
  pages={173--181},
  year={2012},
  publisher={Elsevier}
}

@article{zomorrodi2012optcom,
  title={OptCom: a multi-level optimization framework for the metabolic modeling and analysis of microbial communities},
  author={Zomorrodi, Ali R and Maranas, Costas D},
  journal={PLoS computational biology},
  volume={8},
  number={2},
  pages={e1002363},
  year={2012},
  publisher={Public Library of Science San Francisco, USA}
}

@article{varma1994metabolic,
  title={Metabolic flux balancing: basic concepts, scientific and practical use},
  author={Varma, Amit and Palsson, Bernhard O},
  journal={Bio/technology},
  volume={12},
  number={10},
  pages={994--998},
  year={1994},
  publisher={Nature Publishing Group US New York}
}

@article{bolusani2024scip,
  title={The SCIP optimization suite 9.0},
  author={Bolusani, Suresh and Besan{\c{c}}on, Mathieu and Bestuzheva, Ksenia and Chmiela, Antonia and Dion{\'\i}sio, Jo{\~a}o and Donkiewicz, Tim and van Doornmalen, Jasper and Eifler, Leon and Ghannam, Mohammed and Gleixner, Ambros and others},
  journal={arXiv preprint arXiv:2402.17702},
  year={2024}
}

@article{huangfu2018parallelizing,
  title={Parallelizing the dual revised simplex method},
  author={Huangfu, QHJAJ and Hall, JA Julian},
  journal={Mathematical Programming Computation},
  volume={10},
  number={1},
  pages={119--142},
  year={2018},
  publisher={Springer}
}

@misc{gurobi,
  author = {{Gurobi Optimization, LLC}},
  title = {{Gurobi Optimizer Reference Manual}},
  year = 2026,
  url = "https://www.gurobi.com"
}

\end{document}